\documentclass[a4paper, reprint, twoside, nofootinbib]{revtex4-1}

\usepackage[english]{babel}
\usepackage[utf8]{inputenc}
\usepackage[labelfont=up]{subcaption}
\usepackage[x11names, table]{xcolor}
\usepackage{float}
\usepackage{mathtools}
\usepackage[top=1.in, bottom=1.in, left=0.5in, right=0.5in]{geometry}
\usepackage{amsthm, amssymb}
\usepackage{graphicx}
\usepackage{lipsum}
\usepackage{bm}
\usepackage[pdftex, pdftitle={Article}, pdfauthor={Author}]{hyperref}
\usepackage{wasysym}
\usepackage{enumitem}
\hypersetup{colorlinks=true, citecolor=black, linkcolor=black, urlcolor=blue} 
\usepackage[normalem]{ulem}

\DeclarePairedDelimiter\ton{(}{)}
\DeclarePairedDelimiter\qua{[}{]}

\definecolor{LightRed}{rgb}{1,0.6,0.6}
\definecolor{LightGreen}{rgb}{0.6,1,0.6}

\begin{document}

\title{Quantifying the Unexpected: a scientific approach to Black Swans}

\author{Giordano De Marzo$^{1, 4, 5}$}
\author{Andrea Gabrielli$^{1, 2, 3}$}
\author{Andrea Zaccaria$^{2}$}
\author{Luciano Pietronero$^{1, 2, 4}$}
    
    \affiliation{$^1$Centro Ricerche Enrico Fermi, Piazza del Viminale, 1, I-00184 Rome, Italy.\\
    $^2$Istituto dei Sistemi Complessi (ISC) - CNR, UoS Sapienza,P.le A. Moro, 2, I-00185 Rome, Italy.\\
    $^3$Dipartimento di Ingegneria, Universit\`a degli Studi Roma Tre, Via Vito Volterra 62, I-00146 Rome, Italy\\
    $^4$Dipartimento di Fisica Universit\`a ``Sapienza”, P.le A. Moro, 2, I-00185 Rome, Italy.\\
    $^5$Sapienza School for Advanced Studies, ``Sapienza'', P.le A. Moro, 2, I-00185 Rome, Italy.
    }

\date{\today} 

\begin{abstract} 
	Many natural and socio-economic systems are characterized by power-law distributions that make the occurrence of extreme events not negligible. Such events are sometimes referred to as Black Swans, but a quantitative definition of a Black Swan is still lacking. Here, by leveraging on the properties of Zipf-Mandelbrot law, we investigate the relations between such extreme events and the dynamics of the upper cutoff of the inherent distribution. This approach permits a quantification of extreme events and allows to classify them as White, Grey, or Black Swans. Our criterion is in accordance with some previous findings, but also allows us to spot new examples of Black Swans, such as Lionel Messi and the Turkish Airline Flight 981 disaster. The systematic and quantitative methodology we developed allows a scientific and immediate categorization of rare events, providing also new insight into the generative mechanism behind Black Swans.
\end{abstract}

\maketitle
\section{Introduction}
During the last decade the complexity paradigm \cite{anderson1972more,Jaguar, pietronero2008complexity} has been successfully applied not only to the area of physical systems but also to many other phenomena, including socio-economic systems \cite{castellano2009statistical, tacchella2018dynamical}. \newline  
One of the most ubiquitous features in the complexity science is the emergence of extreme events, orders of magnitude larger than the typical ones. Severe financial crisis, devastating earthquakes, or deadly wars, all can be interpreted in terms of complex inherent structures which give rise to power law distributed phenomena \cite{Newman}. Mandelbrot and Taleb have been among the firsts \cite{mandelbrot2001scaling, taleb2007black} to investigate these phenomena and to stress the limits of standard statistical techniques in such framework, the latter with the introduction of the celebrated Black Swan metaphor. According to Taleb \cite{taleb2007black} a Black Swan event:
	\begin{itemize}
		\item is unexpected and unpredictable;
		\item has a great impact, both positive or negative;
		\item makes people try to explain its occurrence once it has been observed.
	\end{itemize}
If extreme event are found to follow a power law their effects are in some way mitigated because they stop to be totally unexpected. In this case Taleb speaks of Gray Swans \cite{taleb2007black}, which are unpredictable but not unexpected. However, as we are going to show, even if we know that the underlying distribution is power law-like, Black Swans can still occur due to the possible dynamics of the upper cutoff. 
	
The First World War and 9/11 terrorist attacks are often used as examples of Black Swans \cite{taleb2007black, nafday2009strategies, hajikazemi2016black, ferguson2020black}; however, up to now a scientific and quantitative assessment of Black Swans has been lacking and this sometimes allows policy makers to improperly use this terminology. For instance, during the last months many governments, financial institutions and journals identified the COVID-19 pandemic as a Black Swan, even though the possibility of the spreading of a newborn virus had been already pointed out by several studies and even by Taleb himself \cite{taleb2007black}. This arbitrariness is favored by the fact that after the efforts of Taleb and Mandelbrot, most of the discussion about Black Swans is still conducted on a qualitative philosophical/psychological level. 
	
In the following we show, on a quantitative basis, that Black Swans are related to a non-stationarity in the inherent power law distribution, generated by a jump dynamics of the upper cutoff. This implies that when performing risk analysis, the assumption of a time independent probability distribution may result in severe underestimations of extreme events. A central property of Black Swans is that they can not be predicted starting from data about the system under consideration, however, an analysis including also the environment the system is coupled to permits, in some cases, to spot the jumps of the upper cutoff, and so to foresee Black Swans. This new perspective allow us to introduce a quantitative measure of the surprise associated to large events, that we call Blackness, which can be used to scientifically classify them into three distinct sets: White Swans, Grey Swans, and Black Swans. Our quantitative criterion partly confirms what previously stated by Taleb using qualitative arguments, namely that the First World War and 9/11 terrorist attacks have been Black Swans with respect to the number of casualties, while 1987 Black Monday has been only a Grey Swan. New examples of Black Swans we find are the amount of goals scored by Lionel Messi in LaLiga and the number of casualties of Turkish Airline Flight 981 disaster, since both events drastically overcame the previous estimate of the upper cutoff. 
	
	\vspace{1.2em}
	\begin{figure*}[t]
	   	\centering
        \includegraphics[width=0.9\linewidth]{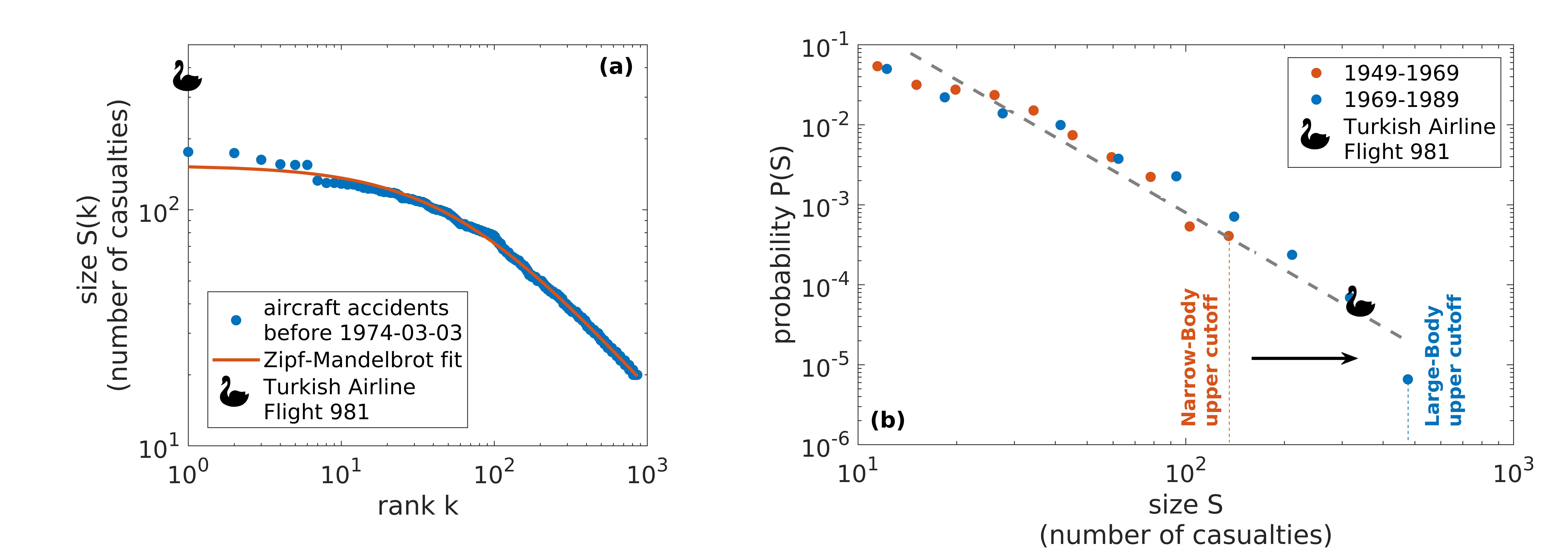}     
	    \caption{\textbf{Black Swans arise from jumps in the upper cutoff.} \textbf{a)} Rank-size plot of airline disasters casualties occurred before 1974-03-03 (blue dots) with the corresponding fit to Zipf-Mandelbrot law (red line). The presence of strong deviations from Zipf's law would have suggested that this set completely sampled the inherent distribution, meaning that the maximal possible number of casualties (i.e., the upper cutoff) was approximately two hundred. The stylized black swan corresponds to Turkish Airline Flight 981 disaster. $346$ people died in that circumstance, subverting the estimate of the upper cutoff. This accident, which has been the first to involve a large-body aircraft, can be then considered a Black Swan. \textbf{b)} Probability distribution of airline disaster casualties. The red dots represent the distribution of accidents occurred between 1949 and 1969, while blue ones show the distribution of those occurred between 1969 and 1989. Approximately the same number of accidents occurred in these two periods ($798$ and $781$, respectively) and so if the distribution would not have changed, then red and blue dots would have spanned the same range of casualties. However, the upper cutoff approximately doubles, making the Black Swan possible.}     
        \label{fig:fig_1_main}
	\end{figure*}
	\section{Results}
	\subsection{Zipf-Mandelbrot law}
	Our scientific framework to analyze Black Swans is given by Zipf's law and, more in general, power laws. Zipf's law \cite{Zipf} is an ubiquitous scaling law found in many natural and socio-economical systems \cite{Newman, blasius2009, furusawa2003, maillart2008, refId0}. Given a system composed of $N$ objects and denoting by $S(k)$ the size of the $k$th largest one, Zipf's law reads
	\begin{equation}	
		S(k)=\frac{S(1)}{k^{\gamma}},
		\label{eq:Zipf}
	\end{equation}
	where $k$ is the rank, $\gamma$ is the Zipf's exponent, and $S(1)$ is the empirical maximum, that is the largest element or event in the system under consideration. 
	Zipf's law is generally visualized by the so called rank-size plot, obtained plotting the ordered sequence of the sizes as function of their position in the sequence; a straight line in loglog scale is thus expected. However, it is common to observe deviations from Zipf's law at low ranks \cite{de2021dynamical, cristelli2012there, burroughs2001}, which can be quantified by introducing a parameter $Q$ in the so called Zipf-Mandelbrot law \cite{Mandelbrot}:
	\begin{equation}	
		S(k)=\frac{\bar{S}}{(k+Q)^{\gamma}}.
		\label{eq:Zipf_Mandelbrot}
	\end{equation}
	The parameter $Q>0$ will play a crucial role in the analysis of Black Swans. As shown in \cite{de2021dynamical}, the Zipf-Mandelbrot law is observed whenever the inherent distribution is a power law and its parameters are related to those of the probability distribution by the following relations 
	\begin{equation}
		\begin{cases}
			\gamma=\frac{1}{\alpha-1}\\
			\bar{S}=N^{\gamma}s_m\\
			Q=N\ton*{\frac{s_m}{s_M}}^{1/\gamma}
		\end{cases}
		\label{eq:gamma_barS_Q}
	\end{equation}
	where 
	\begin{itemize}
		\item $\alpha$ is the exponent of the inherent power law distribution, that is $P(S)\sim S^{-\alpha}$;
		\item $s_m$ and $s_M$ are the lower and upper cutoffs of the distribution, possibly corresponding to physical limits, that is $P(S)=0$ for $S<s_m$ and $S>s_M$;
		\item $N$ is the number of elements in the system or records in the catalog.
	\end{itemize}
	The deviation parameter $Q$ is a measure of the level of sampling \cite{de2021dynamical} (see Methods for details): for $Q\approx0$ the underlying distribution is under-sampled and the upper cutoff can not be inferred, while for $Q\gg1$ it is completely sampled and $s_M$ coincides with the empirical maximum. Indeed by combining Eqs.~\eqref{eq:Zipf_Mandelbrot} and \eqref{eq:gamma_barS_Q} we obtain an expression relating the upper cutoff of the distribution to the measurable parameters $Q$, $S(1)$ and $\gamma$
	\begin{equation}
		s_M=S(1)\ton*{\frac{Q+1}{Q}}^{\gamma}.
		\label{eq:s_M_main}
	\end{equation}
In the limit $Q\to0$ the upper cutoff diverges, meaning that data do not provide a sufficient level of sampling for inferring it. This last expression thus allows, when $Q$ is sufficiently large, to compute the upper cutoff of a power law distribution and will be used in the following to study real systems.
	\vspace{1.5em}

		\begin{figure*}[t]
    	\centering
        \includegraphics[width=\linewidth]{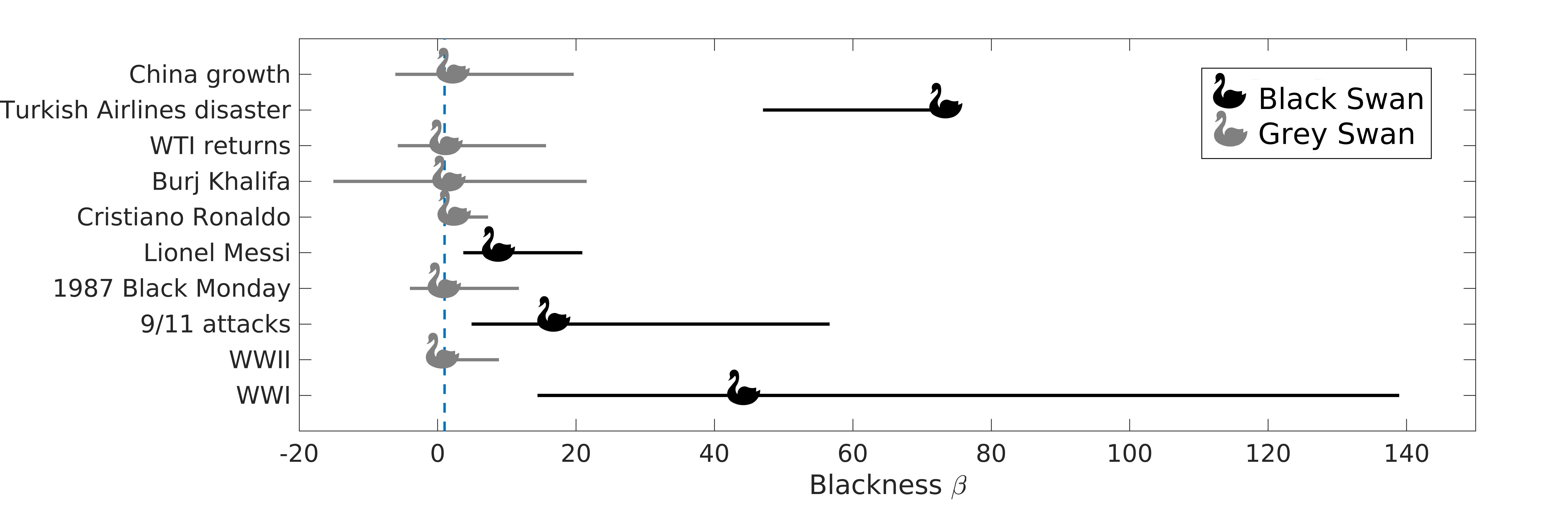}
        \caption{\textbf{Blackness.} Blackness of ten events with $90\%$ confidence bounds. Here the threshold for an event to be considered a Black Swan is $\beta>1$; only four events out of the ten we considered are Black Swans (Turkish Airlines disaster, Lionel Messi, 9/11 terrorist attacks and World War One), while the remaining six are Grey Swans. These results have been obtained by mean of Eq.~\eqref{eq:blackness2}; see Methods for details.}     
        \label{fig:fig_2_main}
	\end{figure*}	
    
    \subsection{The Blackness}
	Let us consider a random number generator extracting values from a power law distribution with unknown parameters $\alpha$, $s_m$ and $s_M$. If we look at the first $N\ll\ton*{s_M/s_m}^{1/\gamma}$ numbers, being $Q\ll1$, the corresponding rank-size plot will be straight \cite{de2021dynamical}. Moreover, our set under-samples the inherent distribution, and so there would be no surprise if the next draw returns a value much larger than those previously observed, because it is not possible to infer the upper cutoff $s_M$. 	
	
	If we keep drawing numbers, sooner or later we will completely sample the inherent distribution, this occurring for $N\gg\ton*{s_M/s_m}^{1/\gamma}$. In this way, the empirical maximum becomes very close to the upper cutoff, that can thus be inferred using Eq.~\eqref{eq:s_M_main}, and nothing unexpected should occur. In this situation Black Swans are absent, but what happens if the upper cutoff increases to $s_M'\gg s_M$? (Note that this actually happens in various social and biological systems, for instance as a consequence of the coupling with an external system or a technological change). The answer is simple, our apparently perfect knowledge of the underlying distribution becomes problematic as soon as a number close to $s_M'$ is extracted. Without knowing if and when the upper cutoff jumps, there is no way not only to predict, but also to expect such an event. This kind of event is what we will classify as a Black Swan.

	This simple example we sketched captures some points that are crucial for understanding the phenomenology of Black Swans (and possibly mitigate their utmost effects):
	\begin{itemize}
		\item an arbitrary large event can be or be not a Black Swan depending on the level of sampling of the distribution, so depending on the value of $Q$. A system showing pure Zipf's law ($Q\approx0$) never gives rise to Black Swans. The idea is that when the level of sampling is low there is no way to characterize the upper limit and so to realize that $s_M \rightarrow s_M'$.
		\item an event can be classified as a Black Swan if and only if it is (much) larger than any event previously observed \textit{and} the deviation parameter $Q$ of the system is large. Indeed, this implies that it is beyond the estimate of the upper limit;
		\item a stationary power law only gives rise to Grey Swans, since when the level of sampling is low there is no surprise if an event much larger than those previously observed occurs, while when the level of sampling is high no event much larger than the observed maximum can occur. Note that for a power law without upper cutoff, the level of sampling is always low, since the support of the distribution is not finite. As a consequence, also in this case, no Black Swan can be observed.
	\end{itemize}
	In order to better clarify these points, we consider a real system showing this kind of dynamics: aircraft accidents. Here the size of the event is given by the number of casualties. In Fig.~\ref{fig:fig_1_main}a we show the corresponding rank size plot computed by considering all events until the 02-03-1974, the day before the crash of the Turkish Airline Flight 981, which is represented by the Black Swan. Strong deviations from Zipf's law are present, and $Q$ is relatively large. On the basis of the previous considerations, one could have concluded that the upper cutoff of the distribution had already been reached, and so no surprise should have been expected. However, the day after, the Flight 981 crashed, provoking the death of 346 people, a number approximately twice as large as the previous most severe accident. This increase was possible because of the introduction in the late sixties of wide-body aircrafts, that can carry approximately twice as many passengers as the older narrow-body models. This implied a sudden increase of the upper cutoff of the casualties distribution, that we depict in Fig.~\ref{fig:fig_1_main}b. In this sense we can conclude that the Flight 981 was indeed a Black Swan. Note that, before and after the jump, the scaling exponent remained the same. \newline
In summary, usually an empirical power law is characterized by three parameters, the exponent and the two cutoffs, and it must be non stationary in order to give rise to Black Swans. Now, the lower cutoff clearly does not play any role regarding large events, while a variation of the exponent influences the frequency of extreme events, but not their size. As a consequence, the only form of non-stationarity that may produce Black Swans is a jump of the upper cutoff.
	\vspace{1.5em}	
		
	\begin{figure*}[t]
    	\centering
        \includegraphics[width=\linewidth]{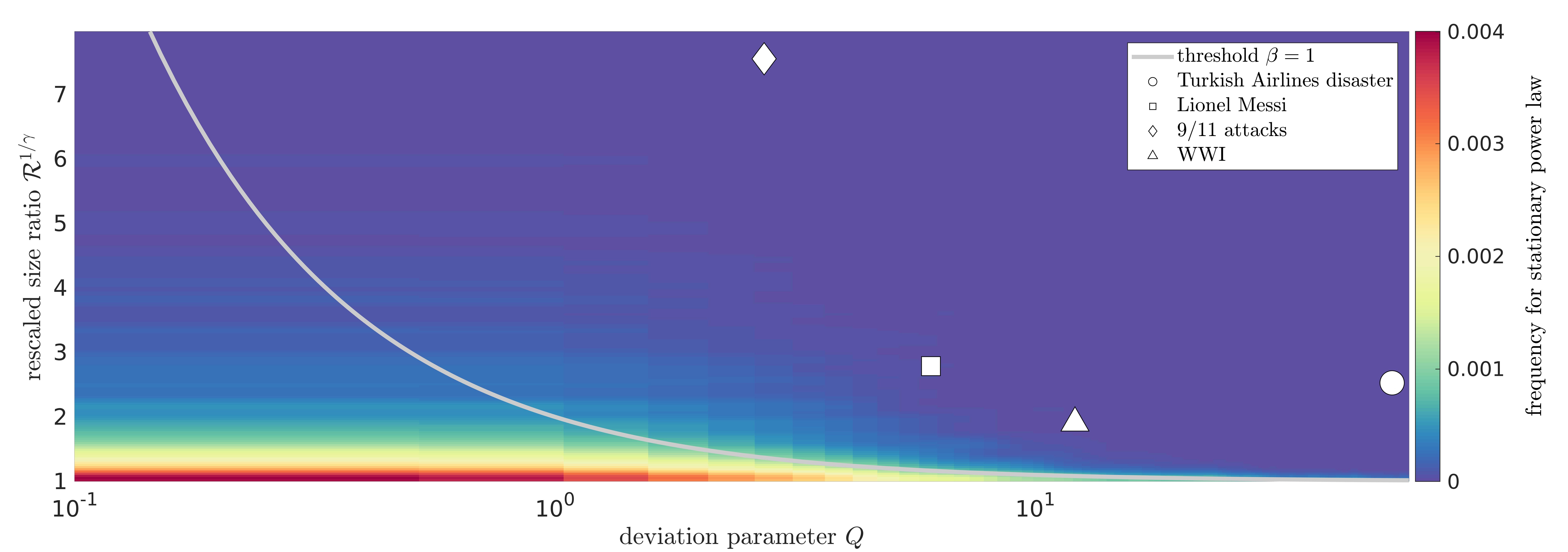}
        \caption{\textbf{Swans Plane.} Visualization in the Swans plane of the outcomes of a truncated stationary power law with $\alpha=1.5$, $s_m=1$ and $s_M=10^6$ (color map) and of the four Black Swans we identified (white dots). Concerning the stationary power law, the figure shows the frequency of events with $\mathcal{R}>1$, obtained exploiting a binning procedure in the Swans plane; lighter colors correspond to higher frequencies. The white curve corresponds to the threshold $\beta=1$ and divides Grey Swans and Black Swans. As expected, the frequency goes to zero above the curve, proving that a stationary power law only produces Grey Swans. Note that as expected all the four Black Swans reside in a region of the Swans Plane where the frequency of power law events is vanishing, so they can not be explained by a stationary power law.}     
        \label{fig:Q_ratio_powerlaw}
	\end{figure*}
	Once the phenomenology of extreme events has been discussed, it is simple to derive a quantitative criterion to classify them in Black, Gray, and White Swans: i) an event smaller than the maximum of those previously occurred is a White Swan; ii) a Gray Swan is an event whose size is larger than those previously observed, but that is not unexpected in the sense that it is lower than the estimated upper cutoff; iii) a Black Swan is an event whose size is larger than those previously observed and that is also unexpected, being larger than the expected upper cutoff.\newline
	We can then define the \textit{Blackness} $\beta$ of a new event of size $S_{new}$ as
	\begin{equation}
		\beta=\frac{S_{new}-S(1)}{s_M-S(1)},
		\label{eq:blackness1}
	\end{equation}
	where $S(1)$ is the empirical maximum and $s_M$ is the best estimate of the upper cutoff we can infer from the data, that is given by Eq.~\eqref{eq:s_M_main}. Formalizing the three definitions just mentioned, $\beta$ can be used to determine the ``color'' of an extreme event:
	\begin{itemize}
		\item \textbf{White Swan:} $S_{new}<S(1) \rightarrow \beta<0$;
		\item \textbf{Grey Swan:} $S(1)<S_{new}<s_M\rightarrow 0<\beta<1$;
		\item \textbf{Black Swan:} $S_{new}>s_M\rightarrow \beta>1$.
 	\end{itemize}
 	As shown in the methods section, $\beta$ can be expressed directly in terms of only empirical quantities
 	\begin{equation}
		\beta=\frac{\mathcal{R}-1}{\ton*{\frac{Q+1}{Q}}^{\gamma}-1},
		\label{eq:blackness2}
	\end{equation}
	where $\mathcal{R}=S_{new}/S(1)$ is the relative size of the new event with respect to the empirical maximum. This last expression shows that $\beta$ well summarizes the two crucial points we stressed above. Namely $\mathcal{R}>1$ is a necessary but not sufficient condition for an event to be a Black Swan, since also the deviation parameter $Q$ (or, equivalently, the level of sampling) must be considered. Indeed if $Q=0$, that is for a perfect Zipf's law, the Blackness goes to zero independently of the relative size. This because in the case of a pure Zipf's law the upper cutoff can not be estimated. \newline
	\vspace{1.5em}	
	
	\subsection{Applications to real systems}
	\subsubsection{Classification of past events}
	The Blackness concept allows to search for Black or Gray swans in any system showing an inherent power law distribution, obtaining an objective and scientific based measure of the ``surprise'' associated to an event. By performing fits to the rank-size plots (see the Method section for details), and by using Eq.~\eqref{eq:blackness2}, we computed the Blackness of a number of extreme events that previous works have found to be power law distributed \cite{clauset2009power, malacarne2000regularities, cirillo2020tail, mandelbrot2001scaling, golyk2012self, chatterjee2017fat} and that range from sport to natural disasters and finance; the results are displayed in Fig.~\ref{fig:fig_2_main}. The ``Blackest'' Black Swan among the events we considered is Turkish Airline disaster, followed by WWI; the size of the latter has been defined as the number of casualties normalized by the world population. For what concerns the WWI, the growing globalization of the world and the extensive usage of new deadly weapons may be the responsible of the jump of the upper cutoff. Surprisingly WWII is only a Grey Swan. Intuitively, the occurrence of the Great War, a conflict much more severe than any previous interstate-war, already proved that the upper cutoff of the distribution had increased, making the second world conflict relatively less surprising in terms of casualties. Another Swan that presents a Blackness lower than one (considering uncertainty) is the 1987 Black Monday. Here the size is given by the module of Dow Jones index daily return; during the Black Monday this index lost about $23\%$ of its value, the worst fall in its history.  Also other ``officially unexpected'' events in finance and economics are found to be Grey Swans: the oil market volatility due to the spreading of Covid-19 (the event we analyzed is the largest monthly fluctuation of West Texas Intermediate (WTI) index (+75\%, occurred between March and April 2020). We considered pre-Covid19 data (1986-2019) and the growth of China (the size is defined as the percentage variation of GDPppp and we considered $35$ years returns of countries between the periods 1865-1900, 1900-1935, and 1950-1985. Countries that grew thanks to oil or natural resources have been excluded, but analogous results are found also considering all the countries). It is interesting to note that according to the common wisdom the growth of China is an incredible outlier \cite{pritchett2014asiaphoria}, while in our framework its growth is a remarkable but not unexpected event, confirming recent analysis based on the methods of analogues \cite{tacchella2018dynamical, cristelli2015heterogeneous}. \newline
	Conversely, the soccer top star Lionel Messi is a Black Swan if measured by the number of goals he scored in LaLiga (455). We can argue that the growth of the upper cutoff is due to the increase of the number of teams taking part in the championships, which, in turn, increased the number of games played in a season and the number of potential goals. Indeed, the previous record holder was Telmo Zarra, who scored 268 goals playing between 1949 and 1957; even if, as Lionel Messi, he scored an average of 0.9 goals per match and played for the same number of seasons (16), the number of teams involved in LaLiga was between 12 and 16, against the 20 of today. Using the same criterion Cristiano Ronaldo, with his $311$ goals, is only a Grey Swan. The September 11 attacks, often used as an archetype of Black Swan events, is characterized by a Blackness much larger than one, confirming the blackness of this terrorist attack in terms of casualties, even if in this case a clear and objective motivation for the jump of the upper cutoff is hard to find. Finally, the Burj Khalifa, that is the tallest building in the world, is only a Grey Swan, despite an height approximately $70\%$ larger than Taipei 101, the previous record holder.\newline				
Setting $\beta=1$ in Eq.~\eqref{eq:blackness2} we can obtain a threshold value $\mathcal{R}_{th}$ for the relative size dividing Black from Grey Swans:
	\begin{equation}
		\mathcal{R}_{th}=\ton*{\frac{Q+1}{Q}}^{\gamma}=\frac{s_M}{S(1)} 
		\label{eq:threshold_R}
	\end{equation}
which once again can expressed in terms of only empirical quantities and coincides with the ratio between the estimate of the upper cutoff and the empirical maximum. 
In this way we can now define the \textit{Swans plane} $Q-\mathcal{R}^{1/\gamma}$ (Fig.~\ref{fig:Q_ratio_powerlaw}), since only these two quantities are needed to determine the nature of an extreme event. Three areas can be identified: the White Swans region ($\mathcal{R}^{1/\gamma}<1$, if the event has a size less than the already seen maximum $S(1)$), the Grey Swans region ($1<\mathcal{R}<\frac{Q+1}{Q}$, a size higher than the maximum, but not surprising given the upper cutoff estimation) and the Black Swans region ($\mathcal{R}>\frac{Q+1}{Q}$, higher than both the maximum and the upper cutoff and so, in this sense, surprising). Note that the trivial White Swan region is not represented in the figure.  \newline
As discussed above, stationary power laws are not expected to produce Black Swans, being such events related to an abrupt non stationarity of the upper cutoff. This is confirmed by Fig.~\ref{fig:Q_ratio_powerlaw}, where we exploited the Swans plane to visualize the outcomes of a stationary truncated power law.	More precisely we generated $N_i=100$ numbers from a stationary truncated power law with parameters $\alpha=1.5$, $s_m=1$ and $s_M=10^6$, so that the initial deviation parameter is $Q_i=0.1$, and then we extracted other $N_f=10^5$ numbers, so to arrive to a final deviations parameter $Q_f=100$. At each draw we checked if the newcomer has been larger than the previously observed numbers. If this was the case, we fitted the sample available before the new extraction occurred with Zipf-Mandelbrot law, determining the value of $Q$; the full procedure has been repeated $10^3$ times. Brighter colors correspond to a larger frequency and, as expected, frequencies go to zero out of the Grey Swan region delimited by the white curve corresponding to $\beta=1$. In the same figure we also plotted the four Black Swans we identified, as it is possible to see they all reside in a region of the Swans Plane where the stationary power law produces no events and so they are not compatible with a stationary power law. \newline
\subsubsection{Future Black Swans}
By using our estimation of the upper cutoff, given by Eq.~\eqref{eq:s_M_main}, we can estimate the size of a newcomer event to be a Black Swan. We computed the upper cutoff of five social and natural systems and, in turn, the minimum size for new events to be Black Swans. We show our results in Table \ref{tab:fig_1_main}. It results that a pandemic should kill more than $5$ billion people - i.e. approximately $40$ times the Black Death, the worst pandemic ever with its $200$ millions deaths - for being considered a Black Swan. This stems from the fact that only few pandemics have been recorded during human history and so the inherent distribution is not much sampled. Conversely, the distributions of wildfires occurred in Alberta (CA) and Italian earthquakes are characterized by an high level of sampling. Indeed a wildfire should be just $1.54$ times larger than the observed maximum\footnote{August 1981 wildfire, $~10^6$ha burned} for being classified as a Black Swan, while an earthquake should release only $1.22$ times the energy of worst Italian earthquake ever\footnote{1693 Val di Noto earthquake, $7.32$Mw}. Finally, the distribution of Interstate Wars and Dow Jones index daily returns are completely undersampled and this reflects in the fact that in these systems any event, no matter how large, can be at most a Grey Swan. More details about these database are reported in the Methods section.
		
		\begin{table}[t]
		\begin{tabular}{ ||c|c|c|| } 
			\hline
			System & $\mathcal{R}_{th}$ & $s_M$ \\
			\hline
			\cellcolor{LightRed} Pandemics & $42.29$ & $5.81\cdot 10^9$ deaths\\
			\cellcolor{LightGreen}Alberta Wildfires &$1.54$& $1.55\cdot10^6$ha\\
			\cellcolor{LightRed} Interstate Wars & $\infty$ & $\infty$\\
			\cellcolor{LightRed} Dow Jones index return & $\infty$ & $\infty$\\
			\cellcolor{LightGreen} Italian Earthquake & $1.23$ & $7.46$M\\
			\hline
		\end{tabular}
		\caption{\textbf{Future Black Swans.} Upper cutoff $s_M$ and relative size $\mathcal{R}_{th}$ of five events that could be considered Black Swans with a confidence of $90\%$. Systems highlighted in red are characterized by a low level of sampling, while those highlighted in green present an high level of sampling.}     
        \label{tab:fig_1_main}
	\end{table}	
	
\section{Discussion}
In this work we exploit the connections among the Zipf-Mandelbrot law, the inherent power law, and the level of sampling, discussed in \cite{de2021dynamical}, to obtain a relation to derive an empirical estimation of the upper cutoff of a power law distribution. We define Black Swans as events whose size is both extreme (i.e., larger than the observed maximum) and unexpected (i.e., larger than the estimated cutoff). We show that three ingredients are needed to produce Black Swans: an inherent truncated power law distribution, an high level of sampling, and a jump dynamics of the upper cutoff. Turkish Airline Flight 981 is a clear-cut example where such a dynamics is particularly evident, since the jump of the upper cutoff has been provoked by a well defined technological advance, the introduction of large body aircrafts. As a consequence, looking only at data regarding a certain system, may result in ignoring the jump of the upper cutoff, not allowing to foresee the possibility of a Black Swan. When performing risk assessment it is instead crucial to analyze also the context the systems lives in, since jumps can be provoked by eternal factors rather than by the laws governing the system itself, and not to rely on the assumption of a stationary probability distribution.\newline 
We also introduced a quantitative and scientific measure of very large events, the Blackness, which allows to divide them in White, Grey, and Black Swans. Using this parameter, which depends on empirical quantities only, we checked that stationary power laws only produce Grey Swans and we analyzed several empirical systems in search of Black Swans. We found out that our criterion is in line with most of the qualitative findings of Taleb, since it correctly classifies the World War I and 9/11 terrorist attacks as Black Swans, while the 1987 Black Monday is classified only as a Grey Swan. We also spotted new examples of Black Swans, such as Turkish Airline disaster and number of goals scored by Lionel Messi. The latter probably connected to an increase of the number of teams involved in LaLiga championship.
Finally, we determined how large should various events be in order to be classified as Black Swan by estimating the upper cutoff of their inherent power law distributions. For instance, in the case of Italian earthquakes the distribution is highly sampled, and so an earthquake releasing just $1.23$ times the energy of the largest earthquake ever recorded would be a Black Swan. Conversely, the distribution of Dow Jones index daily returns is undersampled and so any fluctuation, no matter how much large, would be only an unsurprising Grey Swan. \\
We believe that the introduction of a quantitative criterion to classify extreme events in Black Swans (or not) can be extremely useful not only from a scientific point of view, but also to scientifically ground the discussion among public opinion, academia, and policy makers.

		\begin{acknowledgments}
           Giordano De Marzo is grateful to Federico Attili for useful discussions about the fitting procedure and the computation of confidence bounds.
        \end{acknowledgments}
		\bibliography{bibliography}
	\section*{Methods}
		\subsection*{Analytical results}
			As done in \cite{de2021dynamical}, let us consider a truncated power law distribution of sizes, $P(S)$, that is
			\begin{equation}
			    	P(S)=
				\begin{cases}
					0 \ \text{for} \ S<s_m \\
					\frac{c}{S^{\alpha}} \ \text{for} \ s_m\leq S \leq s_M \\
					0 \ \text{for} \ S>s_M
				\end{cases}
				\label{eq:trunc}
			\end{equation}
			where $c$ is the normalization constant, and $s_{m}$ and $s_{M}$ respectively correspond to the natural lower and upper cutoffs, always present in real systems. These cutoffs are connected to $c$ by the normalization condition
			\begin{equation}
				c\int\limits_{s_m}^{s_M}\frac{ds}{s^{\alpha}}=1 \ \rightarrow \ c=\frac{\alpha-1}{s_m^{1-\alpha}-s_M^{1-\alpha}}
			\label{eq:c}
			\end{equation}
			We can then express the rank-size relation as a function of the PDF parameters using the fact that given the PDF $P(S)$ of a continuous variable $S$, the values of its Cumulative Distribution Function (CDF) $C(S)$, associated to the different values of S, are approximately equiprobable. In fact if $P(s)$ is the PDF of the variable $S$ defined in the interval $[s_m,s_M]$, then $C(S)=\int_{s_m}^S ds'\,P(s')$. By performing the change of variables from $S$ to $C=C(S)$, and calling $f(C)$ its PDF, we get by definition of PDF and CDF $f(C)=\frac{dS(C)}{dC}P(S)|_{S=S(C)}=1$ for $0\le C\le 1$. This implies that, given $N$ values of $S$ independently extracted from $P(S)$, with good approximation they can be taken as uniformly spaced in the corresponding variable $C$. Thus, the $k^{\mbox{\small{th}}}$ size ranked value $S(k)$ approximately corresponds to the CDF value  $\frac{N+1-k}{N+1}$. In formulas
	        \[
	            \int\limits_{s_m}^{S(k)}P(S)dS=c\int\limits_{s_m}^{S(k)}\frac{ds}{s^{\alpha}}\simeq\frac{N+1-k}{N+1}\,,
	        \]
	       which, together to Eq.~\eqref{eq:c}, gives
	        \[
	            \frac{S(k)^{1-\alpha}-s_m^{1-\alpha}}{s_M^{1-\alpha}-s_m^{1-\alpha}}\simeq\frac{N+1-k}{N+1} \,.
		    \]
		    By assuming $N+1\approx N$, $s_M\gg s_m$, and introducing $\gamma=\frac{1}{\alpha-1}$, we end up with the final rank-size formula  
	        \begin{equation*}
	    	    S(k)=\qua*{\frac{Ns_m^{\frac{1}{\gamma}}s_M^{\frac{1}{\gamma}}}{Ns_m^{\frac{1}{\gamma}}+ks_M^{\frac{1}{\gamma}}}}^{\gamma}=\frac{N^{\gamma}s_m}{\qua*{k+N\ton*{\frac{s_m}{s_M}}^{\frac{1}{\gamma}}}^{\gamma}}\,.
	        \end{equation*}
	        By comparing this expression with Zipf-Mandelbrot law, that is
	        \begin{equation}
	         	S(k)=\frac{\bar{S}}{(k+Q)^{\gamma}},
	         	\label{eq:Zipf_Mandelbrot_met}
	        \end{equation}
	        we obtain
	        \begin{equation}
	    	    \begin{cases}
	    	        \gamma=\frac{1}{\alpha-1}\\
	    		    \bar{S}=N^{\gamma}s_m\\
	    		    Q=N\ton*{\frac{s_m}{s_M}}^{\alpha-1}\,
	    	    \end{cases}\label{eq:gamma_barS_Q_met}
	        \end{equation}
			These expressions relate the number of values$/$objects and the parameters of the PDF $P(S)$ on one side, and the Zipf-Mandelbrot parameters on the other. Note that $Q$ not only quantifies deviations from Zipf's law, but also quantifies the level of sampling of the inherent distribution. Indeed $Q$ is:
			\begin{itemize}
				\item the larger the wider is the statistical sample, so the larger is the numerosity of the sample $N$;
				\item the smaller the wider is the extension of the truncated power law, given by the ratio between the upper cutoff and the lower one.
			\end{itemize}		
			
			It is possible to derive an expression connecting the upper cutoff of the distribution to the deviation parameter $Q$, Zipf's exponent $\gamma$ and the empirical maximum $S(1)$. Combining Eqs.~\eqref{eq:Zipf_Mandelbrot_met} and \eqref{eq:gamma_barS_Q_met} we obtain 
			\[
				S(1)=\frac{N^{\gamma}s_m}{(Q+1)^{\gamma}}=s_M\frac{N^{\gamma}\frac{s_m}{s_M}}{(Q+1)^{\gamma}}=s_M\ton*{\frac{Q}{Q+1}}^{\gamma},
			\]
			which yields
			\begin{equation}
				s_M=S(1)\ton*{\frac{Q+1}{Q}}^{\gamma}.
				\label{eq:s_M_met}
			\end{equation}
			As expected for $Q\to\infty$ the empirical maximum $S(1)$ coincides with the upper cutoff and consequently any event larger than $S(1)$ is a surprise, while for $Q\to0$ no event can be surprising since the upper cutoff is diverging. Again we see that $Q$ plays the role of level of sampling quantifier, since the truncation point (the upper cutoff), can be appreciated only if $Q$ is sufficiently large. Exploiting Eq.~\eqref{eq:s_M_met} we can rewrite the Blackness $\beta=(S_{new}-S(1))/(s_M-S(1))$ of an event with size $S_{new}$ as
			\begin{align}
				\beta&=\beta\ton*{\mathcal{R}, Q, \gamma}=\frac{S_{new}-S(1)}{S(1)\qua*{\ton*{\frac{Q+1}{Q}}^{\gamma}-1}}=\nonumber\\
				&=\frac{\mathcal{R}-1}{\ton*{\frac{Q+1}{Q}}^{\gamma}-1},
				\label{eq:surprise_met}
			\end{align}
			where $\mathcal{R}=S_{new}/S(1)$ is the ratio between the size of the new event and the empirical maximum. Since the new event is a Black Swan if its size is larger than the estimate of the upper cutoff the threshold $\mathcal{R}_{th}$ dividing Grey and Black Swans is 
			\begin{equation*}
				\mathcal{R}_{th}=\frac{S_{new}}{s_M}=\ton*{\frac{Q+1}{Q}}^{\gamma}.
			\end{equation*}
			In these terms we can restate the criterion introduced in the main text terms of three empirical quantities: the deviation parameter $Q$, Zipf's exponent $\gamma$ and the ratio $\mathcal{R}=S(1)/s_M$. More precisely 	
			\begin{equation}
				\begin{cases}
					\text{White Swan} \iff \mathcal{R}\leq1\\
					\text{Gray Swan} \iff 1<\mathcal{R}\leq \mathcal{R}_{th}\\
					\text{Black Swan} \iff \mathcal{R}>\mathcal{R}_{th}
				\end{cases},
				\label{eq:criterion_met}
			\end{equation}
			Note that the threshold ratio $\mathcal{R}_{th}=\mathcal{R}_{th}(Q, \gamma)$: i) diverges if $Q$ goes to zero; ii) tends to one for $Q$ going to infinity; iii) is increasing with $\gamma=1/(\alpha-1)$, so it is the larger the fatter is the inherent distribution.
		\subsection*{Databases}
			All the database we used in our analysis are freely accessible on the web, details can be found below.
			\begin{itemize}
				\item \textbf{GDPppp of countries} In order to study the growth of countries so to determine if the growth of China has been a Black Swan, we used Maddison database \cite{maddison2013}, available \href{https://www.rug.nl/ggdc/historicaldevelopment/maddison/releases/maddison-database-2010}{here}, which provides GDP PPP of countries from 1 AD to 2008. We integrated it with IMF \href{https://en.wikipedia.org/wiki/List_of_countries_by_past_and_projected_GDP_(PPP)#cite_note-1}{data} to obtain a database which ranges from 1900 to 2020.
				\item \textbf{Airplane disasters} The analysis of airplanes disasters has been performed exploiting a dataset from Kaggle.com containing data of airplane accidents involving civil, commercial and military transport worldwide from 1908-09-17 to 2009-06-08. Such dataset is available \href{https://www.kaggle.com/saurograndi/airplane-crashes-since-1908}{here}.
				\item \textbf{WTI returns} The monthly returns of West Texas Intermediate Oil can be accessed from different sources, we used a dataset ranging from 1986 to 2020 that can be found \href{https://datahub.io/core/oil-prices}{here}.
				\item \textbf{Tallest buildings} The list of the tallest buildings before Burj Khalifa, so before 2010, has been generated from \url{skyscrapercenter.com}
				\item \textbf{LaLiga top scorers} The list of LaLiga top scorers has been retrieved from \url{transfermarkt.com}. We considered only seasons between 1928/1929 and 2004/2005, so before the blow up of Messi and Ronaldo.
				\item \textbf{Down Jones index returns} Historical daily returns of Dow Jones index from 1986 to 2016 have been downloaded from Quandl.com and can be found \href{https://www.quandl.com/data/BCB/UDJIAD1-Dow-Jones-Industrial-Average}{here}. Data from 2016 to present days have been retrieved for \url{finance.yahoo.com}.
				\item \textbf{Terrorist Attacks} The analysis of casualties provoked by terrorist attacks has been performed using the RAND Database of Worldwide Terrorism Incidents (RDWTI) that is accessible \href{https://www.rand.org/nsrd/projects/terrorism-incidents/download.html}{here}.
				\item \textbf{Interstate Wars} Data on Inter-State wars have been taken from The Correlates of War (COW) Project \cite{sarkees2010}.
				\item \textbf{Pandemics} Regarding pandemics, we exploited the average estimate of the number of casualties described in \cite{cirillo2020tail}.
				\item \textbf{Wildfires} The analysis of wildfires is based on Alberta Wildfire datasets. More precisely we merged the four databases available \href{https://wildfire.alberta.ca/resources/historical-data/historical-wildfire-database.aspx}{here}, obtaining a dataset spanning spanning the period 1961-2018.
				\item \textbf{Italian Earthquakes} For our analysis of the Italian earthquakes we used the INGV Parametric Catalogue of Italian Earthquakes, which "provides homogeneous macroseismic and instrumental data and parameters for Italian earthquakes with maximum intensity $\geq5$ or magnitude $\geq4.0$ in the period 1000-2017" \cite{Earthquakes}.
			\end{itemize}
		\subsection*{Fitting procedure and confidence bounds estimation}
			We adopted a standard non linear least squares (NLS) fitting procedure to determine the parameters of Zipf-Mandelbrot law. The accuracy of this tecnique, when applied to the rank-size plot or to the complementary cumulative distribution, is comparable to maximum likelihood estimates \cite{white2008estimating}, while being much simpler (in the case of unknown upper cutoff). In particular, we used Eq.~\eqref{eq:Zipf_Mandelbrot} partially linearized through logarithms
		\[
			\log S(k)=-\gamma \ln\ton*{k+Q}+c,
		\] 
		where $Q$, $\gamma$ and $c$ are free parameters. All the systems we considered have already been widely studied and there is strong evidence on the presence of an underlying power law distribution, nevertheless we exploited the p-value to check the goodness of our fits. We followed the procedure described in \cite{clauset2009power}, namely:
		\begin{enumerate}
			\item we compute the parameter $Q$ and $\gamma$ of the empirical data with the NLS;
			\item we use Eq.~\eqref{eq:gamma_barS_Q} to determine the parameters of the underlying power law distribution. Note that $N$ and $s_m$ are given, respectively, by the number of elements in the sample and by the size of the smallest object;
			\item we compute the Kolmogorov-Smirnov distance between the empirical data and the power law;
			\item we generate $M=1000$ Monte Carlo samples (each with the same numerosity $N$ of the empirical sample) from the inferred power law distribution and, for each of them, we repeat steps from 1 to 3 so to determine the statistics of the Kolmogorov-Smirnov distance;
			\item the p-value is defined as the fraction of the Monte Carlo samples whose Kolmogorov-Smirnov distance is larger than that of the empirical data.
		\end{enumerate}
		Following \cite{clauset2009power} power law hypothesis is rejected if the p-value is smaller than $0.1$. 
	\begin{table}
		\begin{tabular}{ ||c|c|c|| } 
			\hline
			System & number of elements $N$ & p-value \\
			\hline
			Wars before WWI & $38$ & $0.95$\\
			Wars before WWII & $55$ & $0.30$\\
			Terrorist attacks & $150$ & $0.48$\\
			Dow Jones returns before 1987 & $250$ & $0.24$\\
			LaLiga top scorers & $150$ & $0.59$\\
			Tallest buildings & $500$ & $0.52$\\
			WTI monthly returns & $75$ & $0.66$\\
			Airplanes disasters & $859$ & $0.68$\\
			Countries growth & $70$ & $0.18$\\
			Pandemics & $35$ & $0.89$ \\
			Alberta Wildfires & $250$ & $0.39$ \\
			Wars & $80$ & $0.82$ \\
			Dow Jones returns & $250$ & $0.67$ \\
			Italian Earthquakes & $500$ & $0.19$ \\
			\hline
		\end{tabular}
		\caption{\textbf{p-value.} Number of elements considered and p-value for the systems we analyzed. All the p-values are above the threshold of $0.1$, meaning that all these systems are well described by a power law distribution.}     
        \label{tab:p-value}
	\end{table}
	All the p-values for the systems we analyzed are reported in Tab.~\ref{tab:fig_1_main}, as it is possible to see all values are above the threshold $0.1$, meaning that, as previously noticed, all the systems considered are well described by an underlying power law distribution. Note, however, that in some systems, as in the case of pandemics, only a limited number of elements is available and so the parameters estimated by the fitting procedure may be not very precise. In order to take into account this fact, as explained below, we perform a parametric bootstrap which allows us to include in the uncertainty over the parameters also the effect of the low numerosity.
	
	Once the parameters $Q$ and $\gamma$ have been obtained, the Blackness $\beta$ can be computed exploiting Eq.~\eqref{eq:blackness2}, however determining the uncertainty of such quantity is not trivial. Naively one could propagate on $\beta$ the uncertainty on $Q$ and $\gamma$, $\sigma_Q$ and $\sigma_{\gamma}$ returned by the NLS fitting method. However this methods does not takes into account statistical fluctuations that are encountered considering different samples generated by the same power law distribution. For this reason we exploited a parametric bootstrap so to obtain a more realistic measure of uncertainty. In particular the procedure we adopted is the following
	\begin{itemize}
			\item we compute the parameter $Q$ and $\gamma$ of the empirical data with the NLS and we use them to obtain the Blackness $\beta$;
			\item we use Eq.~\eqref{eq:gamma_barS_Q} to determine the parameters of the underlying power law distribution;
			\item we generate $M=1000$ Monte Carlo samples with numerosity $N$ as the empirical sample using the power law distribution obtained in the previous step;
			\item each synthetic sample $m$ is fitted with the NLS technique, so to obtain the parameters $Q^m$ and $\gamma^m$ and their standard deviation $\sigma_{Q^m}$ and $\sigma_{\gamma^m}$. These quantities are used to determine the Blackness of the event under analysis with respect to the synthetic sample, $\beta^m$, whose uncertainty is obtained propagating $\sigma_{Q^m}$ and $\sigma_{\gamma^m}$ 
			\[
				\sigma_{\beta^m}=\sqrt{\ton*{\frac{d \beta}{d Q}\sigma_{Q^m}}^2+\ton*{\frac{d \beta}{d \gamma}\sigma_{\gamma^m}}^2}.
			\]
			\item the distribution of $\beta$, $P(\beta)$, is obtained as a mixture of $M$ Gaussians with parameters $\beta^m$ and $\sigma_{\beta^m}$
			\[
				P(\beta)=\frac{1}{M}\sum_{m=1}^M \mathcal{N}\ton*{\beta^m, \sigma_{\beta^m}},
			\]
			where $\mathcal{N}\ton*{x, y}$ is a Gaussian with mean $x$ and variance $y^2$;
			\item starting from the probability distribution $P(\beta)$, the confidence bound for $\beta$ is easily obtained using the cumulative distribution and determining the interval containing $90\%$ of the probability.
	\end{itemize}
	
\end{document}